%% file: main-output.tex
\documentclass[adraft]{eptcs}
\providecommand{\event}{MSFP 2012}

\usepackage{breakurl}
\usepackage{amsmath,amssymb,amsthm}
\usepackage{supertabular}
\usepackage{comment}
\usepackage{stmaryrd} 
\usepackage{wasysym} 
\usepackage{xspace}
\usepackage{xcolor}
\usepackage{fancyvrb}

\include{spec}

\renewcommand{\ottkw}{\mathsf}

\newcommand{\gsep}{\;|\;}

\newif\ifcomments\commentsfalse
\ifcomments
\newcommand{\cjc}[1]{\textcolor{blue}{\textbf{[#1 ---CJC]}}}
\newcommand{\vs}[1]{\textcolor{blue}{\textbf{[#1 ---VS]}}}
\newcommand{\scw}[1]{\textcolor{red}{\textbf{[#1 ---SCW]}}}
\else
\newcommand{\cjc}[1]{}
\newcommand{\vs}[1]{}
\newcommand{\scw}[1]{}
\fi

\newcommand{\trellys}{\textsc{Trellys} }

\newtheoremstyle{better}{\parsep}{\topsep}%
     {}
     {}
     {\itshape}
     {.}
     {0.5em}
     {} 
\theoremstyle{better}
\newtheorem{thm}{Theorem}

\newtheorem{lemma}[thm]{Lemma}

\title{Step-Indexed Normalization for a Language with General Recursion}
\author{             Chris Casinghino 
        \qquad\qquad Vilhelm Sj\"{o}berg 
        \qquad\qquad Stephanie Weirich
\institute{School of Engineering and Applied Science\\
University of Pennsylvania}
\email{            ccasin@cis.upenn.edu 
      \qquad\qquad vilhelm@cis.upenn.edu
      \qquad\qquad sweirich@cis.upenn.edu}
}
\def\titlerunning{Step-Indexed Normalization}
\def\authorrunning{C. Casinghino, V. Sj\"{o}berg \& S. Weirich}

\begin{document}
\maketitle

\begin{abstract}
  The \trellys project has produced several designs for practical
  dependently typed languages.  These languages are broken into two
  fragments---a {\em logical} fragment where every term normalizes and
  which is consistent when interpreted as a logic, and a {\em
    programmatic} fragment with general recursion and other convenient
  but unsound features.  In this paper, we present a small example
  language in this style.  Our design allows the programmer to
  explicitly mention and pass information between the two fragments.
  We show that this feature substantially complicates the metatheory
  and present a new technique, combining the traditional Girard--Tait
  method with step-indexed logical relations, which we use to show
  normalization for the logical fragment.
\end{abstract}

\section{Introduction}

The \trellys project is a collaborative initiative to design a
dependently-typed language with simple support for general recursion
and other convenient but logically unsound features.  To this end, the
present authors and their collaborators have proposed languages that
are broken into two fragments: a {\em programmatic} fragment with
support for all the desired language features, and a {\em logical}
fragment which can reason about programs but is itself restricted for
consistency \cite{ccasin:plpv11talk, weirich:mfps-talk, kimmel:plpv12,
  weirich:rta-talk, weirich:dtp-talk}.

As a simple example, consider the following natural number division
function written in a Haskell-like syntax:
\begin{verbatim}
   prog div : Nat -> Nat -> Nat
   prog div n m = if n < m then 0 else 1 + (div (n - m) m)
\end{verbatim}
This function computes the integer division of {\tt n} by {\tt m}
unless {\tt m} is {\tt 0}, in which case it loops forever.  We label
it ``{\tt prog}'' to indicate it must be defined in the programmatic
fragment described above.  Disappointingly, {\tt div} can not be
written directly in popular dependently-typed languages like
Coq~\cite{coq-83} or Agda~\cite{norell:thesis} because it is not
total.

There are many sensible properties a programmer might wish to verify
about {\tt div}.  For example, that {\tt div 6 3} evaluates to {\tt
  2}, or that {\tt div n m <= n} when {\tt m} is not zero.  Even
though {\tt div} itself is in the programmatic fragment, we wish to
state these properties in the consistent logical fragment.  For
example:
\begin{verbatim}
  log div63 : div 6 3 = 2
  log div63 = refl
\end{verbatim}
Above, the program (aka proof) {\tt div63} is tagged with ``{\tt
  log}'' to indicate that it should be typechecked in the logical
fragment. The proof itself is just reflexivity, based on the
operational behavior of {\tt div}.

To encourage incremental verification, such a language should also
include a way for programs which are not known to be terminating to
produce proofs.  For example, programmers implementing a complicated
decision procedure might begin by writing in the programmatic fragment
and come back to prove termination at a later time.  To support
passing the proofs produced by such a procedure to the logical
fragment, the language may include an {\em internalized logicality
  judgement}---programs may assert that other programs typecheck in a
certain fragment.  We use the new type form $\ottnt{A}  \ottsym{@}  \theta$, where
$\theta$ is $\ottkw{L}$ or $\ottkw{P}$ for the logical or programmatic
fragment, to claim that a term has type $\ottnt{A}$ in a particular
fragment.  For example, a SAT solver which is not known to be
terminating might be given the following type:
\begin{verbatim}
  prog solver : (f : Formula) -> Maybe ((Satisfiable f) @ L)
\end{verbatim}
Here, {\tt solver} takes in some representation of a formula and
optionally produces a proof that it is satisfiable.  The type ``{\tt
  (Satisfiable f) @ L}'' indicates that if a proof is produced, it
will typecheck in the logical fragment, even though the procedure
itself is written in the programmatic fragment.

For these internalized judgements to be useful, the language must be
able to produce them in one fragment and use them in another.  In
general, any term which is produced in the logical fragment may be
safely used in the programmatic fragment.  Additionally, values at
certain ``first-order'' types (including $\ottnt{A}  \ottsym{@}  \theta$) may be computed
in the programmatic fragment and safely used by the logical fragment.

The metatheory of languages with this collection of features has
proved challenging.  This paper presents a new technique for
demonstrating the normalization (and thus consistency) of the logical
fragment in such a language.  As we will show, direct adaptations of
the Girard--Tait reducibility
method~\cite{girard:phd,tait:reducibility} are insufficient.  Since
logical terms are permitted to make use of proofs produced
programmatically, it is necessary to simultaneously verify partial
correctness properties of the programmatic fragment.  To this end, our
technique combines the traditional method with step-indexed logical
relations~\cite{ahmed:steps,appel:steps}.

Concretely, our contributions are:
\begin{itemize}
\item A small language with an internalized logicality judgement, sum
  types and recursive types (Section~\ref{sec:language}).  While the
  language is insufficient for our examples, it retains enough
  features to exhibit the difficulties we have encountered with
  traditional proofs (Section~\ref{ssec:traditional}).
\item A new, hybrid technique for proving normalization of the
  language's logical fragment (Section~\ref{ssec:steps}).  This
  technique combines the Girard--Tait reducibility method with a
  step-indexed logical relation for simultaneously verifying partial
  correctness properties of the programmatic fragment.  This
  combination seems to be essential to handle the internalized
  logicality judgement.
\item A formalization of the language's metatheory in Coq, including
  type safety and normalization (Section~\ref{ssec:formalization}).
\item A comparison to related work on dependently-typed languages with
  general recursion and techniques for reasoning about them
  (Section~\ref{sec:related}).
\end{itemize}

The language we consider in this paper is simply-typed and thus
insufficient to represent the examples we have presented so far.
However, this smaller language is still complex enough to exhibit the
difficulties we have encountered in proving normalization, and we are
optimistic that our technique will scale up.

\section{Language Definition}
\label{sec:language}

\begin{figure}
  \[
  \begin{array}{lrl}
    \text{Types} & A,B ::= &
      \ottkw{Unit} \gsep  \ottnt{A} \,{}^{ \theta }\!\! \to   \ottnt{B}  \gsep \ottnt{A}  \ottsym{+}  \ottnt{B} \gsep \ottnt{A}  \ottsym{@}  \theta 
        \gsep \alpha \gsep \mu \, \alpha  \ottsym{.}  \ottnt{A} \smallskip \\
    \text{Terms} & a,b ::= & 
      \ottmv{x} \gsep \ottkw{rec} \, \ottmv{f} \, \ottmv{x}  \ottsym{.}  \ottnt{a} \gsep \ottnt{a} \, \ottnt{b} \gsep \ottkw{box} \, \ottnt{a} 
         \gsep \ottkw{unbox} \, \ottmv{x}  \ottsym{=}  \ottnt{a} \, \ottkw{in} \, \ottnt{b} \\
     & & \hspace{-0.5em}|\  \ottsym{()} \gsep \ottkw{inl} \, \ottnt{a} \gsep \ottkw{inr} \, \ottnt{a}
         \gsep  \ottkw{case}\  \ottnt{a} \ \ottkw{of}\ \{ \ottkw{inl}\  \ottmv{x}  \Rightarrow  \ottnt{a_{{\mathrm{1}}}}  ;
           \ottkw{inr}\  \ottmv{x}  \Rightarrow  \ottnt{a_{{\mathrm{2}}}}  \} 
         \gsep \ottkw{roll} \, \ottnt{a} \gsep \ottkw{unroll} \, \ottnt{a} \smallskip \\
    \text{Language Classifiers} & \theta ::= &
      \ottkw{L} \gsep \ottkw{P} \smallskip\\
    \text{Environments} & \Gamma ::= &  \cdot  \gsep  \Gamma ,  \ottmv{x}  :^{ \theta }  \ottnt{A}  
            \smallskip \\
    \text{Values} & \ottnt{v} ::= & \ottmv{x} \gsep \ottsym{()} \gsep \ottkw{inl} \, \ottnt{v} \gsep \ottkw{inr} \, \ottnt{v}
      \gsep \ottkw{rec} \, \ottmv{f} \, \ottmv{x}  \ottsym{.}  \ottnt{a} \gsep \ottkw{box} \, \ottnt{v} \gsep \ottkw{roll} \, \ottnt{v}
            \smallskip \\
  \end{array}
  \]

  \textit{Syntactic Abbreviation}:
  \[
  \begin{array}{lcll}
    \lambda  \ottmv{x}  \ottsym{.}  \ottnt{a} & \triangleq & \ottkw{rec} \, \ottmv{f} \, \ottmv{x}  \ottsym{.}  \ottnt{a} 
               & \qquad \text{when } \ottmv{f} \notin \mathrm{FV}(a)
  \end{array}
  \]
  \caption{Syntax}
  \label{fig:syntax}
\end{figure}

The language that we consider in this paper is 
a variant of the simply-typed call-by-value lambda calculus with
recursive types and general recursion.  Its syntax is given in
Figure~\ref{fig:syntax}.  The chief novelty is the presence of {\em
  consistency classifiers} $\theta$.  These classifiers are used by
the typing judgement (written $ \Gamma \vdash ^{ \theta } \ottnt{a} : \ottnt{A} $) to divide the
language into two fragments.  The {\em logical fragment}, denoted by
$\ottkw{L}$, is a simply-typed lambda calculus with unit and sums.  As we
will show, all terms in this fragment are normalizing.  The {\em
  programmatic fragment}, denoted by $\ottkw{P}$, adds general recursion
and recursive types.  The programmatic fragment is a strict superset
of the logical fragment: if $ \Gamma \vdash ^{ \ottkw{L} } \ottnt{a} : \ottnt{A} $, then $ \Gamma \vdash ^{ \ottkw{P} } \ottnt{a} : \ottnt{A} $
as well.

Terms in the language may include subexpressions from both
fragments. The $\ottkw{box} \, \ottnt{a}$ term form and corresponding $\ottnt{A}  \ottsym{@}  \theta$
type form mark such transitions.  Intuitively, the judgement $ \Gamma \vdash ^{ \theta } \ottkw{box} \, \ottnt{a} : \ottnt{A}  \ottsym{@}  \theta' $ holds when fragment $\theta$ can safely
observe that $\ottnt{a}$ has type $\ottnt{A}$ in the fragment $\theta'$.

\subsection{The typing judgement}

\begin{figure}
  \framebox{\mbox{$ \Gamma \vdash ^{ \theta } \ottnt{a} : \ottnt{A} $}}
  \[
  \begin{array}{c}
    \ottdruleTVar{} \hspace{.28in} \ottdruleTLam{} \hspace{.28in} \ottdruleTRec{} \\ \\
    \ottdruleTBoxP{} \hspace{.35in} \ottdruleTBoxL{} \hspace{.35in} \ottdruleTBoxLV{} \\ \\
    \ottdruleTUnbox{} \hspace{.33in} \ottdruleTApp{} \hspace{.33in} \ottdruleTUnit{} \\ \\
    \ottdruleTSub{} \hspace{.35in} \ottdruleTFOVal{} \hspace{.35in} \ottdruleTInl{} \\ \\
    \ottdruleTInr{} \hspace{.35in} \ottdruleTCase{} \\ \\
    \ottdruleTRoll{} \hspace{.35in} \ottdruleTUnroll{}
  \end{array}
  \]

  \framebox{\mbox{$\ottkw{FO} \, \ottsym{(}  \ottnt{A}  \ottsym{)}$}}
  \[
  \ottdruleFOUnit{} \hspace{.35in} \ottdruleFOSum{} \hspace{.35in} \ottdruleFOAt{}
  \]

  \caption{Typing Rules}
  \label{fig:typing}
\end{figure}

We now describe the typing rules, given in Figure~\ref{fig:typing}.
As shown in rule \textsc{TVar}, variables in the typing context are
tagged with a fragment.  When a value is substituted for a variable,
the value must check in the corresponding fragment.

The fragments of the language may interact in several ways.  Functions
have arguments that are tagged with consistency classifiers, as in
$ \ottnt{A} \,{}^{ \theta }\!\! \to   \ottnt{B} $.  The $\theta$ here specifies whether the function
must be applied to a logical or programmatic term.  This classifier
does not indicate in which fragment the function itself typechecks,
and functions in each fragment are permitted to take arguments from
the other.  Intuitively, the type may be read as ``$A @ \theta \to
B$'', except that users need not explicitly box up arguments to
functions.  The rules for application (which involve the box form)
ensure this does not cause non-termination in the logical fragment, as
we will discuss shortly.

There are two rules for type-checking functions.  The first,
\textsc{TLam}, checks non-recursive functions in the logical fragment.
Here, $\lambda  \ottmv{x}  \ottsym{.}  \ottnt{b}$ is syntax sugar for $\ottkw{rec} \, \ottmv{f} \, \ottmv{x}  \ottsym{.}  \ottnt{b}$ when $f$ does
not occur free in $b$.  The second rule, \textsc{TRec}, checks
(potentially) recursive functions in the programmatic fragment.
Observe that, in both cases, the consistency classifier for the
argument is carried into the context when checking the body, but does
not directly influence the classifier of the function itself.  The
rules are otherwise standard.

The $\ottkw{box} \, \ottnt{a}$ form effectively internalizes the typing judgement.
It is checked by the three rules, describing the circumstances in
which the fragments may safely talk about each other.  The first rule,
\textsc{TBoxP}, says that the programmatic fragment may internalize
any typing judgement---if $a$ has type $A$ in fragment $\theta$, then
the programmatic fragment can observe that $\ottkw{box} \, \ottnt{a}$ has type
$\ottnt{A}  \ottsym{@}  \theta$.

Rules \textsc{TBoxL} and \textsc{TBoxLV} check \textsf{box} in the
logical fragment and are restricted to ensure termination.  The former
says that if $\ottnt{a}$ itself has type $\ottnt{A}$ in the logical fragment,
then $\ottkw{box} \, \ottnt{a}$ may also be formed in the logical fragment (and
checks at type $\ottnt{A}  \ottsym{@}  \theta$ for any $\theta$, since logical terms are
also programmatic).  The latter permits the logical fragment
to observe that a term checks programmatically.  In that case, the
term must be a value to ensure normalization.  This restriction still
permits the logical fragment to consider programmatic terms (for
example, recursive functions are values).

Rule \textsc{TUnbox} checks the elimination form for boxed terms,
which resembles a ``let''.  The term $\ottkw{unbox} \, \ottmv{x}  \ottsym{=}  \ottnt{a} \, \ottkw{in} \, \ottnt{b}$ typechecks
when $\ottnt{a}$ has type $\ottnt{A}  \ottsym{@}  \theta'$ and $\ottnt{b}$ is parameterized by a
value of type $\ottnt{A}$ in fragment $\theta'$.  Intuitively, $\ottnt{a}$
will be evaluated first, eventually yielding a value $\ottkw{box} \, \ottnt{v}$, and
$\ottnt{v}$ is substituted into $b$.  The operational semantics are
discussed in more detail below.  Note that no additional safety
restrictions occur in this rule---the \textsf{box} introduction rules
handle everything required to ensure that the logical fragment
terminates.

The rule for function application, \textsc{TApp}, makes use of the
infrastructure for internalizing the typing judgement.  Recall that
function types $ \ottnt{A} \,{}^{ \theta }\!\! \to   \ottnt{B} $ demand arguments from a particular
fragment.  The \textsf{box} and $\ottnt{A}  \ottsym{@}  \theta$ constructs already give
us a way to safely check a term in different fragments, so we reuse
them here.  To check the application $\ottnt{a} \, \ottnt{b}$ in the fragment
$\theta$, we check that $\ottnt{a}$ has some function type $ \ottnt{A} \,{}^{ \theta' }\!\! \to   \ottnt{B} $, then check that $\ottkw{box} \, \ottnt{b}$ can be given the type $\ottnt{A}  \ottsym{@}  \theta'$
in the current fragment.  This has the effect of restricting some
applications to programmatic terms in the logical fragment---in
general, programmatic arguments to logical functions must be values,
ensuring termination.

Rules \textsc{TUnit}, \textsc{TInl} and \textsc{TInr} deal with the
introduction forms for the unit and sum base types.  These terms may
be used in either fragment and the typing rules are standard.  Rule
\textsc{TCase} checks the pattern matching elimination form for sums.
Notably, sums that typecheck in one fragment may be eliminated in
another---again we use the \textsf{box} 
infrastructure to ensure that this does not introduce non-termination
into the logical fragment.

Two rules describe the relationship between the fragments.  As already
discussed, any logical term can be used programmatically---this is the
content of rule \textsc{TSub}.  Rule \textsc{TFOVal} is more
surprising.  It allows potentially dangerous programmatic terms to be
used in the logical fragment under certain circumstances.  In
particular, the term must be a value (to ensure termination) and its
type must be ``first order''.  The first-order restriction, formalized
by the $\ottkw{FO} \, \ottsym{(}  \ottnt{A}  \ottsym{)}$ judgement in the same figure, intuitively means
that we move can move {\em data} but not {\em computations} from the
programmatic fragment to the logical one.  For example, moving a
natural number computed in $\ottkw{P}$ to $\ottkw{L}$ is safe, but moving a
function from $\ottkw{P}$ to $\ottkw{L}$ could cause non-termination when the
function is applied.

Importantly, $\ottnt{A}  \ottsym{@}  \theta$ is a first-order type for any $A$.  The
programmatic fragment is permitted to compute logical values,
including logical function values, and pass them back to the logical
fragment.  In a language extended with dependent types, we believe
this would be useful for working with proofs.  For example, a partial
decision procedure could be written in the programmatic fragment and
the resulting proofs could be used in the logical fragment if the
procedure terminates.

Finally, the language includes iso-recursive types~\cite{pierce:tapl}.
These are checked by the two rules \textsc{TRoll} and
\textsc{TUnroll}.  Recursive types are restricted to the programmatic
fragment because they can introduce non-termination.

\subsection{Operational Semantics}

\begin{figure}
  \[
  \begin{array}{lrl}
    \text{Evaluation contexts} & \mathcal{E} ::= &
      [\cdot] \gsep [\cdot]  \ottnt{b} \gsep \ottnt{v}  [\cdot] \gsep \ottkw{inl} \, [\cdot] \gsep \ottkw{inr} \, [\cdot]
          \gsep  \ottkw{case}\  [\cdot] \ \ottkw{of}\ \{ \ottkw{inl}\  \ottmv{x}  \Rightarrow  \ottnt{a_{{\mathrm{1}}}}  ;
           \ottkw{inr}\  \ottmv{x}  \Rightarrow  \ottnt{a_{{\mathrm{2}}}}  \} 
            \smallskip \\
      & & \hspace{-0.5em} |\  \ottkw{box} \, [\cdot] \gsep \ottkw{unbox} \, \ottmv{x}  \ottsym{=}  [\cdot] \, \ottkw{in} \, \ottnt{a}
          \gsep \ottkw{roll} \, [\cdot] \gsep \ottkw{unroll} \, [\cdot]
  \end{array}
  \]

  \framebox{\mbox{$\ottnt{a}  \leadsto  \ottnt{b}$}}
  \[
  \begin{array}{c}
    \ottdruleSBeta{} \qquad \ottdruleSUnbox{} \\ \\
    \ottdruleSCaseL{} \qquad \ottdruleSUnroll{} \\ \\
    \ottdruleSCaseR{} \qquad \ottdruleSCtx{}
  \end{array}
  \]

  \[
  \begin{array}{ll}
    \framebox{\mbox{$ \ottnt{a}   \leadsto ^{ \ottnt{n} }  \ottnt{b} $}}&
    \framebox{\mbox{$ \ottnt{a}   \leadsto ^{\ast}  \ottnt{b} $}} \smallskip\\

    \begin{array}{c}
      \ottdruleMSRefl{} \qquad \ottdruleMSStep{}
    \end{array} \qquad \qquad &
    \begin{array}{c}
      \ottdruleASAny{}
    \end{array}
  \end{array}
  \]
  \caption{Operational Semantics}
  \label{fig:step}
\end{figure}

The language's operational semantics are given in
Figure~\ref{fig:step}.  We use standard call-by-value evaluation
contexts and a small-step reduction relation.  Note that reduction
occurs inside $\ottkw{box} \, \ottnt{a}$ terms, motivating some of the
restrictions from the previous section.  The multi-step reduction
relation is indexed by a natural number---this will be useful in the
step-indexed logical relation defined in Section~\ref{sec:metatheory}.

\section{Metatheory}
\label{sec:metatheory}

We now consider the metatheory of the small language presented in
Section~\ref{sec:language}.  Of particular interest is the
normalization result for the logical fragment, for which we employ a
novel combination of traditional and step-indexed logical relations.
We motivate and explain this technique in
Section~\ref{ssec:steps}.  The system also enjoys standard
type-safety properties, as we show in Section~\ref{ssec:type-safety}.
All the results presented here have been mechanized using the Coq
theorem prover, and we briefly describe the formalization in
Section~\ref{ssec:formalization}.  For this reason, we focus on a
high-level description of the techniques and elide most proofs.

\subsection{Type Safety}
\label{ssec:type-safety}

We prove type safety via syntactic progress and preservation
theorems~\cite{wright:syntactic}.  The progress result is direct by
induction on typing derivations, using appropriate canonical forms
lemmas.

\begin{thm}[Progress]
  If $  \cdot  \vdash ^{ \theta } \ottnt{a} : \ottnt{A} $ then either $a$ is a value or $\ottnt{a}  \leadsto  \ottnt{a'}$
  for some $\ottnt{a'}$.
\end{thm}

For preservation, a substitution lemma is required.  Because variables
are values and our language includes a value restriction (in the
\textsc{TBoxLV} rule), we prove the substitution lemma only for
values.

\begin{lemma}[Substitution]
  If $  \Gamma ,  \ottmv{x}  :^{ \theta' }  \ottnt{B}  \vdash ^{ \theta } \ottnt{a} : \ottnt{A} $ and $ \Gamma \vdash ^{ \theta' } \ottnt{v} : \ottnt{B} $, then 
  $ \Gamma \vdash ^{ \theta } \ottsym{[}  \ottnt{v}  \ottsym{/}  \ottmv{x}  \ottsym{]}  \ottnt{a} : \ottnt{A} $.
\end{lemma}

Since we employ a call-by-value operational semantics, this
substitution lemma is enough to prove preservation.

\begin{thm}[Preservation]
  If $ \Gamma \vdash ^{ \theta } \ottnt{a} : \ottnt{A} $ and $\ottnt{a}  \leadsto  \ottnt{a'}$, then $ \Gamma \vdash ^{ \theta } \ottnt{a'} : \ottnt{A} $.
\end{thm}

\subsection{Adapting the Girard-Tait Reducibility Method}
\label{ssec:traditional}

To motivate the use of step-indexed logical relations in our
normalization proof, we will first revisit the standard Girard--Tait
reducibility method~\cite{girard:phd,tait:reducibility} and examine
why more direct adaptations of it fail.  Traditional techniques for
proving strong normalization typically begin by defining the
``interpretation'' of each type.  That is, for each type $A$, a set of
terms $\llbracket A \rrbracket$ is defined approximating the type $A$
and where each term in the set is known to be strongly normalizing.
Then a ``soundness'' theorem is proved, demonstrating that if $a$ has
type $A$ then $a \in \llbracket A \rrbracket$.  This implies $a$ is
strongly normalizing.

\subsubsection{First attempt: ignoring the programmatic fragment}
\label{sssec:firsttry}

We begin by modify this technique in two ways to fit our setting.
First, since we have a deterministic call-by-value operational
semantics, the interpretation of each type will be a set of values
(not arbitrary terms).  Second, since the terms at a given type differ
in the programmatic and logical fragments, we index the
interpretation by $\theta$, writing $ \llbracket  \ottnt{A}  \rrbracket^{ \theta } $.

It is tempting to think that, because we do not care about the
normalization behavior of the programmatic fragment, the programmatic
interpretation of types can be very simple.  Perhaps, for example,
just the well-typed values of the appropriate type will do.  Consider
the following interpretation:
\[
\begin{array}{lcl}
 \llbracket  \ottnt{A}  \rrbracket^{ \ottkw{P} }  &=& \{ v\ |\   \cdot  \vdash ^{ \ottkw{P} } \ottnt{v} : \ottnt{A}  \}
\\
\\
 \llbracket  \ottkw{Unit}  \rrbracket^{ \ottkw{L} }   &=& \{ () \}
\\
 \llbracket  \ottnt{A}  \ottsym{+}  \ottnt{B}  \rrbracket^{ \ottkw{L} }  &=& \{ \ottkw{inl} \, \ottnt{v} \gsep v \in  \llbracket  \ottnt{A}  \rrbracket^{ \ottkw{L} }  \} \cup 
                   \{ \ottkw{inr} \, \ottnt{v} \gsep v \in  \llbracket  \ottnt{B}  \rrbracket^{ \ottkw{L} }  \}
\\
 \llbracket   \ottnt{A} \,{}^{ \theta }\!\! \to   \ottnt{B}   \rrbracket^{ \ottkw{L} }  &=& \{ \lambda  \ottmv{x}  \ottsym{.}  \ottnt{a}\ |\  
                    \cdot  \vdash ^{ \ottkw{L} } \lambda  \ottmv{x}  \ottsym{.}  \ottnt{a} :  \ottnt{A} \,{}^{ \theta }\!\! \to   \ottnt{B}  
    \mbox{ and for any } v \in  \llbracket  \ottnt{A}  \rrbracket^{ \theta } ,
            \ottsym{[}  \ottnt{v}  \ottsym{/}  \ottmv{x}  \ottsym{]}  \ottnt{a}   \leadsto ^{\ast}  \ottnt{v'}  \in  \llbracket  \ottnt{B}  \rrbracket^{ \ottkw{L} }  \}
\\
 \llbracket  \ottnt{A}  \ottsym{@}  \theta  \rrbracket^{ \ottkw{L} }  &=& \{ \ottkw{box} \, \ottnt{v} \ |\ v \in  \llbracket  \ottnt{A}  \rrbracket^{ \theta }  \}
\\
 \llbracket  \mu \, \alpha  \ottsym{.}  \ottnt{A}  \rrbracket^{ \ottkw{L} } &=& \emptyset
\\
 \llbracket  \alpha  \rrbracket^{ \ottkw{L} }  &=& \emptyset
\end{array}
\]
Here, the logical interpretation of $\ottkw{Unit}$ contains only $\ottsym{()}$.
The logical interpretation of a sum type $\ottnt{A}  \ottsym{+}  \ottnt{B}$ contains $\ottkw{inl} \, \ottnt{v}$ for every $v$ in the interpretation $\ottnt{A}$, and $\ottkw{inr} \, \ottnt{v}$ for
every $\ottnt{v}$ in the interpretation of $\ottnt{B}$.  The logical
interpretation of functions types is standard, except for the addition
of the consistency classifier: $ \ottnt{A} \,{}^{ \theta }\!\! \to   \ottnt{B} $ contains the term
$\lambda  \ottmv{x}  \ottsym{.}  \ottnt{a}$ if, for any $\ottnt{v}$ in the interpretation of the domain,
$\ottsym{[}  \ottnt{v}  \ottsym{/}  \ottmv{x}  \ottsym{]}  \ottnt{a}$ reduces to a value in the interpretation of the range
(``related functions take related arguments to related results'').
The logical interpretation of $\ottnt{A}  \ottsym{@}  \theta$ comprises the values $\ottkw{box} \, \ottnt{v}$ where $\ottnt{v}$ is in $ \llbracket  \ottnt{A}  \rrbracket^{ \theta } $.  Finally, the logical
interpretations of recursive types and type variables are empty, since
these are used only in the programmatic fragment.

Before we can state a soundness theorem, we must account for contexts.
We use $\rho$ for mappings of variables to terms, and write $ \Gamma   \models   \rho $ if $ \ottmv{x}  :^{ \theta }  \ottnt{A} \in \Gamma $ implies $ \rho   \ottmv{x}  \in  \llbracket  \ottnt{A}  \rrbracket^{ \theta } $.
We let $ \rho   \ottnt{a} $ stand for the simultaneous replacement of the
variables in $\ottnt{a}$ by the corresponding terms in $\rho$.

In this setting, we would hope to be able to prove the following
soundness theorem:
\begin{quote}
  {\bf Soundness} (take 1): Suppose $ \Gamma \vdash ^{ \ottkw{L} } \ottnt{a} : \ottnt{A} $ and $ \Gamma   \models   \rho $.  Then $  \rho   \ottnt{a}    \leadsto ^{\ast}  \ottnt{v}  \in  \llbracket  \ottnt{A}  \rrbracket^{ \ottkw{L} } $.
\end{quote}
In a proof by induction on the typing derivation, most of the cases
offer little resistance (the interested reader is encouraged to write
out the case for the \textsc{TLam} and \textsc{TApp} rules).  However,
the proof gets stuck at the case for the first order rule:
\[
\ottdruleTFOVal{}
\]
Here, we must show that $ \rho   \ottnt{v}  \in  \llbracket  \ottnt{A}  \rrbracket^{ \ottkw{L} } $ (substituting values
into a value produces a value, so $ \rho   \ottnt{v} $ does not step).
However, since the premise is in the programmatic fragment, we have no
induction hypothesis for $v$.  If $\ottnt{A} = \ottkw{Unit}$, we can complete
the case using a canonical forms lemma (since we know by a
substitution lemma that $  \cdot  \vdash ^{ \ottkw{L} }  \rho   \ottnt{v}  : \ottkw{Unit} $).  However, if
$\ottnt{A}$ is $\ottnt{B}  \ottsym{@}  \ottkw{L}$ we are stuck.  We could use a canonical forms
lemma to observe that $ \rho   \ottnt{v} $ must have the shape $\ottkw{box} \, \ottnt{v'}$,
but no induction hypothesis for $\ottnt{v'}$ is available.

\subsubsection{Second attempt: partial correctness for the
  programmatic fragment}

Our previous attempt failed because the language permits values of
first-order types to move from the programmatic fragment to the
logical fragment, but the theorem we were trying to prove didn't
capture any information about the programmatic fragment.  To fix this,
we might try making two changes.  First, the programmatic and logical
interpretations should agree at first-order types.  Second, the
programmatic interpretation and the soundness theorem should be
modified to prove a partial correctness result for the programmatic
fragment---we'll need to know that {\em if} a programmatic term
normalizes, {\em then} it is in the appropriate interpretation.

These changes should allow us to handle the previously problematic
\textsc{TFOVal} case.  Consider the following modified interpretation,
ignoring recursive types for the moment:
\[
\begin{array}{lcl}
 \llbracket  \ottkw{Unit}  \rrbracket^{ \theta }   &=& \{ () \}
\\
 \llbracket  \ottnt{A}  \ottsym{+}  \ottnt{B}  \rrbracket^{ \theta }  &=& \{ \ottkw{inl} \, \ottnt{v} \gsep v \in  \llbracket  \ottnt{A}  \rrbracket^{ \theta }  \} \cup 
                    \{ \ottkw{inr} \, \ottnt{v} \gsep v \in  \llbracket  \ottnt{B}  \rrbracket^{ \theta }  \}
\\
 \llbracket   \ottnt{A} \,{}^{ \theta }\!\! \to   \ottnt{B}   \rrbracket^{ \ottkw{L} }  &=& \{ \lambda  \ottmv{x}  \ottsym{.}  \ottnt{a}\ |\  
                    \cdot  \vdash ^{ \ottkw{L} } \lambda  \ottmv{x}  \ottsym{.}  \ottnt{a} :  \ottnt{A} \,{}^{ \theta }\!\! \to   \ottnt{B}  
    \mbox{ and for any } v \in  \llbracket  \ottnt{A}  \rrbracket^{ \theta } ,
            \ottsym{[}  \ottnt{v}  \ottsym{/}  \ottmv{x}  \ottsym{]}  \ottnt{a}   \leadsto ^{\ast}  \ottnt{v'}  \in  \llbracket  \ottnt{B}  \rrbracket^{ \ottkw{L} }  \}
\\
 \llbracket   \ottnt{A} \,{}^{ \theta }\!\! \to   \ottnt{B}   \rrbracket^{ \ottkw{P} }  &=& \{ \ottkw{rec} \, \ottmv{f} \, \ottmv{x}  \ottsym{.}  \ottnt{a}\ |\  
                    \cdot  \vdash ^{ \ottkw{P} } \ottkw{rec} \, \ottmv{f} \, \ottmv{x}  \ottsym{.}  \ottnt{a} :  \ottnt{A} \,{}^{ \theta }\!\! \to   \ottnt{B}   \\
    & & \phantom{ \{ \ottkw{rec} \, \ottmv{f} \, \ottmv{x}  \ottsym{.}  \ottnt{a}\ |}
       \mbox{ and for any } v \in  \llbracket  \ottnt{A}  \rrbracket^{ \theta } ,
       \mbox{ if }  \ottsym{[}  \ottnt{v}  \ottsym{/}  \ottmv{x}  \ottsym{]}  \ottsym{[}  \ottkw{rec} \, \ottmv{f} \, \ottmv{x}  \ottsym{.}  \ottnt{a}  \ottsym{/}  \ottmv{f}  \ottsym{]}  \ottnt{a}   \leadsto ^{\ast}  \ottnt{v'} 
       \mbox{ then } \ottnt{v'} \in  \llbracket  \ottnt{B}  \rrbracket^{ \ottkw{P} }  \}
\\
 \llbracket  \ottnt{A}  \ottsym{@}  \theta'  \rrbracket^{ \theta }  &=& \{ \ottkw{box} \, \ottnt{v} \ |\ v \in  \llbracket  \ottnt{A}  \rrbracket^{ \theta' }  \}
\end{array}
\]
Here, the logical interpretation is unchanged.  The programmatic
interpretation of the first-order types is now the same as the logical
interpretation.  Finally, we have modified the programmatic
interpretation of function types to state a partial correctness
property: {\em if} a function terminates when passed a value in the
interpretation of its domain, {\em then} the result must be in the
interpretation of its range.  We now restate the soundness theorem
similarly.
\begin{quote}
  {\bf Soundness} (take 2): Suppose $ \Gamma \vdash ^{ \theta } \ottnt{a} : \ottnt{A} $ and $ \Gamma   \models   \rho $.
  \begin{itemize}
  \item If $\theta$ is $\ottkw{L}$, then $  \rho   \ottnt{a}    \leadsto ^{\ast}  \ottnt{v}  \in  \llbracket  \ottnt{A}  \rrbracket^{ \ottkw{L} } $.
  \item If $\theta$ is $\ottkw{P}$ and $  \rho   \ottnt{a}    \leadsto ^{\ast}  \ottnt{v} $, then $v \in  \llbracket  \ottnt{A}  \rrbracket^{ \ottkw{P} } $.
  \end{itemize}
\end{quote}
With the modified interpretation and soundness theorem, the
\textsc{TFOVal} case now goes through.  Because the rule only applies
to values, the theorem now yields a useful induction hypothesis for
the premise.

Unfortunately, this style of definition introduces a new problem: the
programmatic interpretation of recursive types.  The previous
definition (from Section~\ref{sssec:firsttry}) is insufficient to
handle the \textsc{TUnroll} case of the new soundness theorem.  To
extend our partial correctness property, we might demand that when
unrolling results in a value, that value is in the interpretation of
the unrolled type:
\[
\begin{array}{lcl}
   \llbracket  \mu \, \alpha  \ottsym{.}  \ottnt{A}  \rrbracket^{ \ottkw{P} }  
     &=& \{ \ottkw{roll} \, \ottnt{v} \ |\   \cdot  \vdash ^{ \ottkw{P} } \ottkw{roll} \, \ottnt{v} : \mu \, \alpha  \ottsym{.}  \ottnt{A}  
          \mbox { and } v \in  \llbracket  \ottsym{[}  \mu \, \alpha  \ottsym{.}  \ottnt{A}  \ottsym{/}  \alpha  \ottsym{]}  \ottnt{A}  \rrbracket^{ \ottkw{P} }  \}
\end{array}
\]
However, this is not a valid definition.  If the interpretation is a
function defined by recursion on the structure of types, the
substitution in $ \llbracket  \ottsym{[}  \mu \, \alpha  \ottsym{.}  \ottnt{A}  \ottsym{/}  \alpha  \ottsym{]}  \ottnt{A}  \rrbracket^{ \ottkw{P} } $ ruins its
well-foundedness.

\subsection{A step-indexed interpretation}
\label{ssec:steps}

Happily, a technique exists in the literature to cope with the
circularity introduced by iso-recursive types.  Step-indexed logical
relations~\cite{ahmed:steps,appel:steps} add an index to the
interpretation, indicating the number of available future execution
steps.  Terms in the relation are guaranteed to respect the property
in question only for the number of steps indicated.  The
interpretation is defined recursively on this additional index,
circumventing the circularity problem we encountered above.

Step-indexed logical relations intuitively describe
partial-correctness properties---terms are certified to be well
behaved for a finite number of steps.  For this reason, they have
typically been used to prove safety and program equivalence
properties, not normalization.  We will adopt a hybrid approach, where
the indices track execution of subterms in the programmatic fragment
(where we need a partial correctness result) but not in the logical
fragment (for which we are proving normalization).

Following Ahmed~\cite{ahmed:steps}, our interpretation is split into
two parts.  The {\em value} interpretation $ \mathcal{V}[\![  \ottnt{A}  ]\!]^{ \theta }_{ \ottnt{k} } $ resembles
the interpretations shown in the previous sections.  The $k$ index
here indicates that when a value appears in a larger term, its
programmatic components will be ``well-behaved'' for at least $k$
steps of computation.  The {\em computational} interpretation $ \mathcal{C}[\![  \ottnt{A}  ]\!]^{ \theta }_{ \ottnt{k} } $ contains closed terms, not just values.  Its definition
resembles the statement of the soundness theorem from the previous
section, with steps counted explicitly.  Terms in $ \mathcal{C}[\![  \ottnt{A}  ]\!]^{ \ottkw{L} }_{ \ottnt{k} } $
are guaranteed to normalize to values in $ \mathcal{V}[\![  \ottnt{A}  ]\!]^{ \ottkw{L} }_{ \ottnt{k} } $.  On the
other hand, we have a partial correctness property for terms in $ \mathcal{C}[\![  \ottnt{A}  ]\!]^{ \ottkw{P} }_{ \ottnt{k} } $---{\em if} they reach a value in $j$ steps for some $j \leq
k$, {\em then} the value is in $ \mathcal{V}[\![  \ottnt{A}  ]\!]^{ \ottkw{P} }_{  \ottnt{k}  -  \ottnt{j}  } $.
\[
\begin{array}{lcl}
 \mathcal{V}[\![  \ottkw{Unit}  ]\!]^{ \theta }_{ \ottnt{k} }   &=& \{ () \}
\\
 \mathcal{V}[\![  \ottnt{A}  \ottsym{+}  \ottnt{B}  ]\!]^{ \theta }_{ \ottnt{k} }   &=& \{ \ottkw{inl} \, \ottnt{v} \gsep v \in  \mathcal{V}[\![  \ottnt{A}  ]\!]^{ \theta }_{ \ottnt{k} }  \} \cup 
                         \{ \ottkw{inr} \, \ottnt{v} \gsep v \in  \mathcal{V}[\![  \ottnt{B}  ]\!]^{ \theta }_{ \ottnt{k} }  \}

\\
 \mathcal{V}[\![  \ottnt{A}  \ottsym{@}  \theta'  ]\!]^{ \theta }_{ \ottnt{k} }  &=& \{ \ottkw{box} \, \ottnt{v} \ |\ v \in  \mathcal{V}[\![  \ottnt{A}  ]\!]^{ \theta' }_{ \ottnt{k} }  \}
\\
 \mathcal{V}[\![   \ottnt{A} \,{}^{ \theta' }\!\! \to   \ottnt{B}   ]\!]^{ \ottkw{L} }_{ \ottnt{k} }  &=& \{ \ottkw{rec} \, \ottmv{f} \, \ottmv{x}  \ottsym{.}  \ottnt{a}\ |\  
                    \cdot  \vdash ^{ \ottkw{L} } \ottkw{rec} \, \ottmv{f} \, \ottmv{x}  \ottsym{.}  \ottnt{a} :  \ottnt{A} \,{}^{ \theta' }\!\! \to   \ottnt{B}  \\
    && \phantom{\{ \ottkw{rec} \, \ottmv{f} \, \ottmv{x}  \ottsym{.}  \ottnt{a}}
    \mbox { and } \forall j \leq k, \mbox{ if } v \in  \mathcal{V}[\![  \ottnt{A}  ]\!]^{ \theta' }_{ \ottnt{j} } 
                      \mbox{ then } \ottsym{[}  \ottnt{v}  \ottsym{/}  \ottmv{x}  \ottsym{]}  \ottnt{a} \in  \mathcal{C}[\![  \ottnt{B}  ]\!]^{ \ottkw{L} }_{ \ottnt{j} }  \}
\\
 \mathcal{V}[\![   \ottnt{A} \,{}^{ \theta' }\!\! \to   \ottnt{B}   ]\!]^{ \ottkw{P} }_{ \ottnt{k} }  &=& \{ \ottkw{rec} \, \ottmv{f} \, \ottmv{x}  \ottsym{.}  \ottnt{a}\ |\  
                    \cdot  \vdash ^{ \ottkw{P} } \ottkw{rec} \, \ottmv{f} \, \ottmv{x}  \ottsym{.}  \ottnt{a} :  \ottnt{A} \,{}^{ \theta' }\!\! \to   \ottnt{B}   \\
    && \phantom{\{ \ottkw{rec} \, \ottmv{f} \, \ottmv{x}  \ottsym{.}  \ottnt{a}}
    \mbox { and } \forall j < k, \mbox{ if } v \in  \mathcal{V}[\![  \ottnt{A}  ]\!]^{ \theta' }_{ \ottnt{j} }  
                      \mbox{ then } \ottsym{[}  \ottnt{v}  \ottsym{/}  \ottmv{x}  \ottsym{]}  \ottsym{[}  \ottkw{rec} \, \ottmv{f} \, \ottmv{x}  \ottsym{.}  \ottnt{a}  \ottsym{/}  \ottmv{f}  \ottsym{]}  \ottnt{a}
                                        \in  \mathcal{C}[\![  \ottnt{B}  ]\!]^{ \ottkw{P} }_{ \ottnt{j} } 
 \}
\\
 \mathcal{V}[\![  \mu \, \alpha  \ottsym{.}  \ottnt{A}  ]\!]^{ \ottkw{L} }_{ \ottnt{k} } &=& \emptyset 
\\
 \mathcal{V}[\![  \mu \, \alpha  \ottsym{.}  \ottnt{A}  ]\!]^{ \ottkw{P} }_{ \ottnt{k} } &=& \{ \ottkw{roll} \, \ottnt{v} \ |\ 
       \cdot  \vdash ^{ \ottkw{P} } \ottkw{roll} \, \ottnt{v} : \mu \, \alpha  \ottsym{.}  \ottnt{A}  \mbox{ and } \forall j < k, 
     v \in  \mathcal{V}[\![  \ottsym{[}  \mu \, \alpha  \ottsym{.}  \ottnt{A}  \ottsym{/}  \alpha  \ottsym{]}  \ottnt{A}  ]\!]^{ \ottkw{P} }_{ \ottnt{j} }  \}
\\
\\
\\
 \mathcal{C}[\![  \ottnt{A}  ]\!]^{ \ottkw{P} }_{ \ottnt{k} }  &=& \{ a \gsep   \cdot  \vdash ^{ \ottkw{P} } \ottnt{a} : \ottnt{A}  \mbox{ and }
                   \forall j \leq k, \mbox{ if }  \ottnt{a}   \leadsto ^{ \ottnt{j} }  \ottnt{v}  
                   \mbox{ then } v \in  \mathcal{V}[\![  \ottnt{A}  ]\!]^{ \ottkw{P} }_{ \ottsym{(}   \ottnt{k}  -  \ottnt{j}   \ottsym{)} }  \}
\\
 \mathcal{C}[\![  \ottnt{A}  ]\!]^{ \ottkw{L} }_{ \ottnt{k} }  &=& \{ a \gsep   \cdot  \vdash ^{ \ottkw{L} } \ottnt{a} : \ottnt{A}  
  \mbox{ and }    \ottnt{a}   \leadsto ^{\ast}  \ottnt{v}  \in  \mathcal{V}[\![  \ottnt{A}  ]\!]^{ \ottkw{L} }_{ \ottnt{k} }  \} 
\end{array}
\]

The value interpretation is similar to the proposed interpretation in
the previous section, with two changes.  First, the function type
cases now refer to the computation interpretation rather than
explicitly mentioning the reduction behavior.  Second, the step
indices track reductions in the programmatic fragment.  In particular,
note that the programmatic interpretation of function types demands
that related functions take related arguments to related results at
all {\em strictly smaller} indices, effectively counting the one beta
reduction step that this definition unfolds.  The beta step in the
logical interpretation is not counted, since we are tracking only the
reduction of programmatic components.

Unlike the proposed definition from the previous section, this
interpretation is well defined.  We can formalize its descending
well-founded metric as a lexicographically ordered triple
$(k,A,\mathcal{I})$: here, $k$ is the index, $A$ is the type and
$\mathcal{I}$ is one of $\mathcal{C}$ or $\mathcal{V}$ with
$\mathcal{V} < \mathcal{C}$.  The third element of the triple
tracks which interpretation is being called---the computational
interpretation may call the value interpretation at the same index and
type, but not vice-versa.

\subsection{Normalization}

The step-indexed interpretation from the previous section repairs the
problems encountered in the first two proposed interpretations and can
be used to prove normalization for the logical fragment.  Since our
results are formalized in Coq, we give only a high-level overview of
the proof here.  To begin, we must update the $ \Gamma   \models   \rho $ judgement
to account for steps.  We now write $ \Gamma \models _{ \ottnt{k} }  \rho $ when $ \ottmv{x}  :^{ \theta }  \ottnt{A} \in \Gamma $ implies $ \rho   \ottmv{x}  \in  \mathcal{V}[\![  \ottnt{A}  ]\!]^{ \theta }_{ \ottnt{k} } $.

Three key lemmas are needed in the main soundness theorem.  The first
is a standard ``downward closure'' property that often accompanies
step-indexed logical relations.  This lemma captures the idea that we
build a more precise interpretation of a type by considering terms
that must be valid for more steps.
\begin{quote}
  {\bf Lemma} (Downward Closure):  For any $A$ and $\theta$, if $j
  \leq k$ then $ \mathcal{V}[\![  \ottnt{A}  ]\!]^{ \theta }_{ \ottnt{k} } \subseteq  \mathcal{V}[\![  \ottnt{A}  ]\!]^{ \theta }_{ \ottnt{j} } $ and $ \mathcal{C}[\![  \ottnt{A}  ]\!]^{ \theta }_{ \ottnt{k} }  \subseteq  \mathcal{C}[\![  \ottnt{A}  ]\!]^{ \theta }_{ \ottnt{j} } $.
\end{quote}
We have two lemmas relating the programmatic and logical
interpretations, corresponding to the \textsc{TFOVal} and
\textsc{TSub} typing rules.  The first says that the two
interpretations agree on first-order types:
\begin{quote}
  {\bf Lemma}: If $\ottkw{FO} \, \ottsym{(}  \ottnt{A}  \ottsym{)}$, then $ \mathcal{V}[\![  \ottnt{A}  ]\!]^{ \ottkw{L} }_{ \ottnt{k} }  =  \mathcal{V}[\![  \ottnt{A}  ]\!]^{ \ottkw{P} }_{ \ottnt{k} } $.
\end{quote}
The second captures the idea that the logical fragment is a subsystem
of the programmatic fragment:
\begin{quote}
  {\bf Lemma}: For any $A$ and $k$, $ \mathcal{V}[\![  \ottnt{A}  ]\!]^{ \ottkw{L} }_{ \ottnt{k} }  \subseteq  \mathcal{V}[\![  \ottnt{A}  ]\!]^{ \ottkw{P} }_{ \ottnt{k} } $ and $ \mathcal{C}[\![  \ottnt{A}  ]\!]^{ \ottkw{L} }_{ \ottnt{k} }  \subseteq  \mathcal{C}[\![  \ottnt{A}  ]\!]^{ \ottkw{P} }_{ \ottnt{k} } $.
\end{quote}

The content of the soundness theorem is essentially the same as in our
second failed attempt, but we can now state it more directly, using
the computational interpretation.  The theorem is proved by induction
on the typing derivation, using the lemmas outlined above.
\begin{quote}
  {\bf Theorem} (Soundness): If $ \Gamma \vdash ^{ \theta } \ottnt{a} : \ottnt{A} $ and $ \Gamma \models _{ \ottnt{k} }  \rho $, then $ \rho   \ottnt{a}  \in  \mathcal{C}[\![  \ottnt{A}  ]\!]^{ \theta }_{ \ottnt{k} } $.
\end{quote}
The normalization of the logical fragment is a direct consequence of
this theorem and the definition of the interpretation.
\begin{quote}
  {\bf Lemma} (Normalization): If $  \cdot  \vdash ^{ \ottkw{L} } \ottnt{a} : \ottnt{A} $ then there exists
  a value $v$ such that $ \ottnt{a}   \leadsto ^{\ast}  \ottnt{v} $.
\end{quote}

\subsection{Formalization}
\label{ssec:formalization}

The proof outline above has been formalized with the Coq proof
assistant~\cite{coq-83}.  The proof scripts are written in a heavily
automated style, inspired by Chlipala's work on practical dependently
typed programming~\cite{chlipala:cpdt,chlipala:cpdtjfp}.  They are
available for download at the first author's website:
\begin{quote}
  \url{http://www.seas.upenn.edu/~ccasin/papers/step_normalization.tar.gz}.
\end{quote}
The language formalized differs in several minor ways from the one
presented in this paper.  Namely,
\begin{itemize}
\item de Bruijn indices are used for binding instead of explicit names.
\item Rather than being syntactic sugar, $\lambda  \ottmv{x}  \ottsym{.}  \ottnt{a}$ is a separate form
  in the grammar of expressions.
\item The reduction relation is formalized with explicit congruence
  rules rather than evaluation contexts.
\item The formalized language includes natural numbers, but not unit.
\end{itemize}
Additionally, to prove certain facts about the interpretation, we
found it necessary to add a standard axiom of functional
extensionality to Coq.  This axiom is known to be consistent
with Coq's logic~\cite{coqfaq}.

\section{Related Work}
\label{sec:related}

\subsection{Step-indexed logical relations}

Our proof technique draws heavily from previous work on step-indexed
logical relations.  The idea to approximate models of programming
languages up to a number of remaining execution steps originated in
the work of Appel and McAllester on foundational proof-carrying
code~\cite{appel:steps}.  They observed that the step indices allowed
a natural interpretation of recursive types.  Subsequently, Ahmed
extended this technique to languages involving impredicative
polymorphism, mutable state and other
features~\cite{ahmed:steps,ahmed:phd}.

Hobor, Dockins and Appel have proposed a general {\em theory of
  indirection} which captures many of the common use-cases for
step-indexed models~\cite{hobor09:knot}.  They provide a general
framework for applying these approximation techniques to resolve
certain types of apparent circularity (similar to the problems with
recursive types described above).  In a recent
draft~\cite{dockins:terminationviaindirection}, Dockins and Hobor have
used this framework to provide a Hoare logic of total correctness for
a small language with function pointers and semantic assertions.  This
work is closely related to the present development, but with different
goals: they prove the soundness of a logic which can reason about
termination, while we prove that every term in the logical fragment of
our language terminates.  We have not yet investigated whether their
framework can be adapted to our setting, but this connection is a
promising avenue for future work.

\subsection{Other approaches to recursion and partiality}

Many authors have considered language features to model partiality and
recursion in a consistent dependent type theory.  The language
described in the present paper is much simpler, but our goal is to
provide a foundation from which we may scale up to full dependent
types, so we compare with some of the most closely related approaches.

\paragraph{Partiality monad}
Capretta proposed representing potentially non-terminating
computations via a coinductive {\em partiality
  monad}~\cite{capretta:general}.  This technique can be used in
existing languages like Coq and Agda, which already support
coinduction~\cite{coquand:infinite}.  For example, Agda's partiality
monad has been used to present subtyping for recursive
types~\cite{nils:subtyping} and represent potentially infinite parsing
trees~\cite{nils:parsers}.

There are several differences between these approaches and the one
outlined in this paper.  Coinduction is a very general method for
representing infinite data, which we do not consider.  Our approach
has the advantage that terminating and potentially partial functions
are defined and reasoned about in the same way.  By contrast, the
reasoning principles for coinductively defined functions in Coq and
Agda require the user to consider so-called {\em guardedness
conditions} that are not present for terminating functions.  More, we
are optimistic that splitting the language into two fragments will
allow us to include various other potential sources of logical
unsoundness uniformly, restricting them to the programmatic fragment.
Admittedly, it remains to be seen how well this will work in practice
and whether our proof technique will scale.

\paragraph{Partial Types}
Constable and Smith~\cite{constable:partial-objects} proposed adding
partiality to the Nuprl type theory through the addition of a type
$\overline{A}$ of potentially nonterminating computations of type $A$.
The general fixpoint operator, for defining recursive computations
then has type
\[ 
  (\overline{A} \rightarrow \overline{A}) \rightarrow \overline{A}.
\]
However, to preserve the consistency of the logic in dependent type
theories, the type $A$ must be restricted to {\em admissible}.
types. Crary~\cite{crary:phd} provides an expressive axiomatization of
admissible types, but the resulting conditions lead to significant
proof obligations, in particular when using $\Sigma$ types. Although
we have not yet formally proven the soundness of the system with
arbitrary dependent types (including $\Sigma$ types), we do not
believe that there will be any restrictions on the {\em programmatic}
language, similar to admissibility.

\paragraph{The ``later'' modality}
Nakano~\cite{Nakano:modality} introduced the ``later'' modality to
define a total language with guarded recursive types.  Intuitively, a
term of type $\bullet A$ (pronounced ``later $A$'') will be useable as
a term of type $A$ in the future.  The recursive type $\mu \, \alpha  \ottsym{.}  \ottnt{A}$
then unfolds to $[\bullet \mu \, \alpha  \ottsym{.}  \ottnt{A}/\alpha]A$ rather than $\ottsym{[}  \mu \, \alpha  \ottsym{.}  \ottnt{A}  \ottsym{/}  \alpha  \ottsym{]}  \ottnt{A}$.  Using this modality, he is able to give the type
$(\bullet A \to A) \to A$ to the Y combinator.  This type allows
programmers to define a variety of recursive functions while still
ensuring that the language is normalizing.  Nakano uses a step-indexed
realizability interpretation to prove the normalization property for
his langauge, suggesting deep connections with the present work.  One
substantial difference is that Nakano's calculus is not call-by-value.

The later modality has been used by subsequent authors to design
langauges for a variety of purposes.  Krishnaswami and Benton use it
define a total language for functional reactive
programming~\cite{krishnaswami:ultrametric,krishnaswami:semantic}.
Birkedal et al.~\cite{birkedal:synthetic} study the topos of trees,
which they observe can model an extension of Nakano's calculus to a
full dependent type theory with guarded recursion.  While these
authors do not consider languages with partiality and their settings
have substantial differences from our own, their success in extending
step-indexing and closely-related techniques to model recursion in
larger languages is promising.

\paragraph{Other \trellys approaches}
The \trellys group has been working simultaneously on an alternative
design, where the logical and programmatic languages are completely
separate at a syntactic level~\cite{kimmel:plpv12}.  This considerably
simplifies the metatheory for the logical language, which is no longer
a general programming language but rather a collection of principles
for reasoning about the programmatic language.  On the other hand, it
can restrict the expressiveness of the logic and create duplication
between the two fragments.  We are exploring these trade-offs in our
ongoing research.

\subsection{Modal type systems for distributed computation}

Modal logics allow one to reason from multiple perspectives, called
``possible worlds''.  It is tempting to view the language presented
here as such a system, where the possible worlds are $\theta$, the
logical and computational fragments of the language.

One way to define a modal logic is to make the world explicit, for
example using a judgement $\Gamma \vdash^\theta A$, stating that
under the assumptions in $\Gamma$, the proposition $A$ is true at the
world $\theta$.  Each assumption in the context is tagged with the
world where it holds $(\theta, A)$. 
\[
\frac{
(\theta,A) \in \Gamma
}{
\Gamma \vdash^\theta A
}
\]

In such as system, the \textsf{at} modality~\cite{Jia04modalproofs},
internalizes the typing judgment into a proposition, with introduction
form
\[
\frac{
  \Gamma \vdash^{\theta'} A \qquad
}{
  \Gamma \vdash^{\theta} A @ \theta'
}
\]
and elimination form:
\[
\frac{
  \Gamma \vdash^{\theta} A @ \theta' \qquad
  \Gamma, (\theta',A) \vdash^{\theta} C
}{
 \Gamma \vdash^{\theta} C
}
\]

Our first-order rule is similar to the perspective-shifting rule,
called {\bf get}, from ML5~\cite{murphy2007ml5,murphy2008modal}.
\[
\frac{
  \Gamma \vdash^{\theta'} A   \quad
  A\ \mathsf{mobile}
}{
  \Gamma \vdash^\theta  A
}
\]
This rule, shown above, allows a class of propositions to be directly
translated between worlds.  The class of mobile types is very similar
to our class of first-order types. For example, base types (such as
strings and integers) and the \textsf{at} modality ($\ottnt{A}  \ottsym{@}  \theta$) are
always mobile, sums ($A + B$) are mobile only when their components
($A$ and $B$) are mobile, but implications are never mobile.

One difference between our system and modal logics is our treatment of
implication (i.e. function types).  The functions in our system are
annotated with a domain fragment, but this is not typically the case
in modal logics, where the domain and range of implications are in the
same world.  Such an approach is incompatible with our subsumption
rule:
\[
  \ottdruleTSub{}
\]
Suppose $A$ were a function type $\ottnt{B_{{\mathrm{1}}}} \to \ottnt{B_{{\mathrm{2}}}}$ with no tag on
the domain.  When we defined such a function in the logical fragment,
the function's body could make use of the fact that its argument
checks logically.  If the subsumption rule were used to transport the
function to the programmatic fragment, it could be applied to terms
that check only programmatically, potentially violating assumptions of
its body.

\section{Conclusion}

In this paper, we have presented a small language with two fragments.
The {\em programmatic} fragment supports general recursion and
recursive types, while every term in the {\em logical} fragment is
normalizing.  Despite these differences, each fragment may explicitly
mention and manipulate terms from the other using the {\em
  internalized logicality type}, $\ottnt{A}  \ottsym{@}  \theta$.  We showed that direct
adaptations of the Girard--Tait reducibility method fail to yield a
normalization proof for the logical fragment.  Finally, we proposed a
new technique involving step-indexed logical relations and used it to
complete the proof, which has been formalized in Coq.

The language considered here is small and unsuitable for real
programming tasks.  However, it constitutes the core of one of our
designs for the \trellys programming language, and the metatheoretic
difficulties we explained and solved in this paper also appear there.
In future work, we plan to add polymorphism, type-level computation
and dependent types back to this system.  If our proof technique
scales, this will provide the basis for a practical, dependently-typed
programming language which can naturally express and reason about
non-terminating computations.

\bibliographystyle{eptcs}
\bibliography{refs}
\end{document}

%% file: spec.tex
\newcommand{\ottdrule}[4][]{{\displaystyle\frac{\begin{array}{l}#2\end{array}}{#3}\quad\ottdrulename{#4}}}

\newcommand{\ottpremise}[1]{ #1 \\}
\newenvironment{ottdefnblock}[3][]{ \framebox{\mbox{#2}} \quad #3 \\[0pt]}{}

\newcommand{\ottnt}[1]{\mathit{#1}}
\newcommand{\ottmv}[1]{\mathit{#1}}
\newcommand{\ottkw}[1]{\mathbf{#1}}
\newcommand{\ottsym}[1]{#1}

\newcommand{\ottdrulename}[1]{\textsc{#1}}

\newcommand{\ottdruleSCaseL}[1]{\ottdrule[#1]{%
}{
 \ottkw{case}\  \ottkw{inl} \, \ottnt{v} \ \ottkw{of}\ \{ \ottkw{inl}\  \ottmv{x}  \Rightarrow  \ottnt{a_{{\mathrm{1}}}}  ;
           \ottkw{inr}\  \ottmv{x}  \Rightarrow  \ottnt{a_{{\mathrm{2}}}}  \}   \leadsto  \ottsym{[}  \ottnt{v}  \ottsym{/}  \ottmv{x}  \ottsym{]}  \ottnt{a_{{\mathrm{1}}}}}{%
{\ottdrulename{SCaseL}}{}%
}}

\newcommand{\ottdruleSCaseR}[1]{\ottdrule[#1]{%
}{
 \ottkw{case}\  \ottkw{inr} \, \ottnt{v} \ \ottkw{of}\ \{ \ottkw{inl}\  \ottmv{x}  \Rightarrow  \ottnt{a_{{\mathrm{1}}}}  ;
           \ottkw{inr}\  \ottmv{x}  \Rightarrow  \ottnt{a_{{\mathrm{2}}}}  \}   \leadsto  \ottsym{[}  \ottnt{v}  \ottsym{/}  \ottmv{x}  \ottsym{]}  \ottnt{a_{{\mathrm{2}}}}}{%
{\ottdrulename{SCaseR}}{}%
}}

\newcommand{\ottdruleSBeta}[1]{\ottdrule[#1]{%
}{
\ottsym{(}  \ottkw{rec} \, \ottmv{f} \, \ottmv{x}  \ottsym{.}  \ottnt{a}  \ottsym{)} \, \ottnt{v}  \leadsto  \ottsym{[}  \ottnt{v}  \ottsym{/}  \ottmv{x}  \ottsym{]}  \ottsym{[}  \ottkw{rec} \, \ottmv{f} \, \ottmv{x}  \ottsym{.}  \ottnt{a}  \ottsym{/}  \ottmv{f}  \ottsym{]}  \ottnt{a}}{%
{\ottdrulename{SBeta}}{}%
}}

\newcommand{\ottdruleSUnbox}[1]{\ottdrule[#1]{%
}{
\ottkw{unbox} \, \ottmv{x}  \ottsym{=}  \ottkw{box} \, \ottnt{v} \, \ottkw{in} \, \ottnt{b}  \leadsto  \ottsym{[}  \ottnt{v}  \ottsym{/}  \ottmv{x}  \ottsym{]}  \ottnt{b}}{%
{\ottdrulename{SUnbox}}{}%
}}

\newcommand{\ottdruleSUnroll}[1]{\ottdrule[#1]{%
}{
\ottkw{unroll} \, \ottsym{(}  \ottkw{roll} \, \ottnt{v}  \ottsym{)}  \leadsto  \ottnt{v}}{%
{\ottdrulename{SUnroll}}{}%
}}

\newcommand{\ottdruleSCtx}[1]{\ottdrule[#1]{%
\ottpremise{\ottnt{a}  \leadsto  \ottnt{b}}%
}{
\mathcal{E}  \ottsym{[}  \ottnt{a}  \ottsym{]}  \leadsto  \mathcal{E}  \ottsym{[}  \ottnt{b}  \ottsym{]}}{%
{\ottdrulename{SCtx}}{}%
}}

\newcommand{\ottdruleMSRefl}[1]{\ottdrule[#1]{%
}{
 \ottnt{a}   \leadsto ^{ \ottsym{0} }  \ottnt{a} }{%
{\ottdrulename{MSRefl}}{}%
}}

\newcommand{\ottdruleMSStep}[1]{\ottdrule[#1]{%
\ottpremise{ \ottnt{a}   \leadsto ^{ \ottnt{k} }  \ottnt{b}  \quad \ottnt{b}  \leadsto  \ottnt{b'}}%
}{
 \ottnt{a}   \leadsto ^{ \ottsym{(}   \ottnt{k}  +  \ottsym{1}   \ottsym{)} }  \ottnt{b'} }{%
{\ottdrulename{MSStep}}{}%
}}

\newcommand{\ottdruleASAny}[1]{\ottdrule[#1]{%
\ottpremise{ \ottnt{a}   \leadsto ^{ \ottnt{k} }  \ottnt{b} }%
}{
 \ottnt{a}   \leadsto ^{\ast}  \ottnt{b} }{%
{\ottdrulename{ASAny}}{}%
}}

\newcommand{\ottdruleFOUnit}[1]{\ottdrule[#1]{%
}{
\ottkw{FO} \, \ottsym{(}  \ottkw{Unit}  \ottsym{)}}{%
{\ottdrulename{FOUnit}}{}%
}}

\newcommand{\ottdruleFOSum}[1]{\ottdrule[#1]{%
\ottpremise{\ottkw{FO} \, \ottsym{(}  \ottnt{A}  \ottsym{)} \quad \ottkw{FO} \, \ottsym{(}  \ottnt{B}  \ottsym{)}}%
}{
\ottkw{FO} \, \ottsym{(}  \ottnt{A}  \ottsym{+}  \ottnt{B}  \ottsym{)}}{%
{\ottdrulename{FOSum}}{}%
}}

\newcommand{\ottdruleFOAt}[1]{\ottdrule[#1]{%
}{
\ottkw{FO} \, \ottsym{(}  \ottnt{A}  \ottsym{@}  \theta  \ottsym{)}}{%
{\ottdrulename{FOAt}}{}%
}}

\newcommand{\ottdruleTVar}[1]{\ottdrule[#1]{%
\ottpremise{ \ottmv{x}  :^{ \theta }  \ottnt{A} \in \Gamma }%
}{
 \Gamma \vdash ^{ \theta } \ottmv{x} : \ottnt{A} }{%
{\ottdrulename{TVar}}{}%
}}

\newcommand{\ottdruleTApp}[1]{\ottdrule[#1]{%
\ottpremise{ \Gamma \vdash ^{ \theta } \ottnt{a} :  \ottnt{A} \,{}^{ \theta' }\!\! \to   \ottnt{B}  }%
\ottpremise{ \Gamma \vdash ^{ \theta } \ottkw{box} \, \ottnt{b} : \ottnt{A}  \ottsym{@}  \theta' }%
}{
 \Gamma \vdash ^{ \theta } \ottnt{a} \, \ottnt{b} : \ottnt{B} }{%
{\ottdrulename{TApp}}{}%
}}

\newcommand{\ottdruleTLam}[1]{\ottdrule[#1]{%
\ottpremise{  \Gamma ,  \ottmv{x}  :^{ \theta }  \ottnt{A}  \vdash ^{ \ottkw{L} } \ottnt{b} : \ottnt{B} }%
}{
 \Gamma \vdash ^{ \ottkw{L} } \lambda  \ottmv{x}  \ottsym{.}  \ottnt{b} :  \ottnt{A} \,{}^{ \theta }\!\! \to   \ottnt{B}  }{%
{\ottdrulename{TLam}}{}%
}}

\newcommand{\ottdruleTRec}[1]{\ottdrule[#1]{%
\ottpremise{   \Gamma ,  \ottmv{y}  :^{ \theta }  \ottnt{A}  ,  \ottmv{f}  :^{ \ottkw{P} }   \ottnt{A} \,{}^{ \theta }\!\! \to   \ottnt{B}   \vdash ^{ \ottkw{P} } \ottnt{a} : \ottnt{B} }%
}{
 \Gamma \vdash ^{ \ottkw{P} } \ottkw{rec} \, \ottmv{f} \, \ottmv{y}  \ottsym{.}  \ottnt{a} :  \ottnt{A} \,{}^{ \theta }\!\! \to   \ottnt{B}  }{%
{\ottdrulename{TRec}}{}%
}}

\newcommand{\ottdruleTSub}[1]{\ottdrule[#1]{%
\ottpremise{ \Gamma \vdash ^{ \ottkw{L} } \ottnt{a} : \ottnt{A} }%
}{
 \Gamma \vdash ^{ \ottkw{P} } \ottnt{a} : \ottnt{A} }{%
{\ottdrulename{TSub}}{}%
}}

\newcommand{\ottdruleTFOVal}[1]{\ottdrule[#1]{%
\ottpremise{ \Gamma \vdash ^{ \ottkw{P} } \ottnt{v} : \ottnt{A}  \quad \ottkw{FO} \, \ottsym{(}  \ottnt{A}  \ottsym{)}}%
}{
 \Gamma \vdash ^{ \ottkw{L} } \ottnt{v} : \ottnt{A} }{%
{\ottdrulename{TFOVal}}{}%
}}

\newcommand{\ottdruleTUnit}[1]{\ottdrule[#1]{%
}{
 \Gamma \vdash ^{ \theta } \ottsym{()} : \ottkw{Unit} }{%
{\ottdrulename{TUnit}}{}%
}}

\newcommand{\ottdruleTInl}[1]{\ottdrule[#1]{%
\ottpremise{ \Gamma \vdash ^{ \theta } \ottnt{a} : \ottnt{A} }%
}{
 \Gamma \vdash ^{ \theta } \ottkw{inl} \, \ottnt{a} : \ottnt{A}  \ottsym{+}  \ottnt{B} }{%
{\ottdrulename{TInl}}{}%
}}

\newcommand{\ottdruleTInr}[1]{\ottdrule[#1]{%
\ottpremise{ \Gamma \vdash ^{ \theta } \ottnt{b} : \ottnt{B} }%
}{
 \Gamma \vdash ^{ \theta } \ottkw{inr} \, \ottnt{b} : \ottnt{A}  \ottsym{+}  \ottnt{B} }{%
{\ottdrulename{TInr}}{}%
}}

\newcommand{\ottdruleTCase}[1]{\ottdrule[#1]{%
\ottpremise{ \Gamma \vdash ^{ \theta } \ottkw{box} \, \ottnt{a} : \ottsym{(}  \ottnt{A_{{\mathrm{1}}}}  \ottsym{+}  \ottnt{A_{{\mathrm{2}}}}  \ottsym{)}  \ottsym{@}  \theta' }%
\ottpremise{  \Gamma ,  \ottmv{x}  :^{ \theta' }  \ottnt{A_{{\mathrm{1}}}}  \vdash ^{ \theta } \ottnt{b_{{\mathrm{1}}}} : \ottnt{B}  \quad   \Gamma ,  \ottmv{x}  :^{ \theta' }  \ottnt{A_{{\mathrm{2}}}}  \vdash ^{ \theta } \ottnt{b_{{\mathrm{2}}}} : \ottnt{B} }%
}{
 \Gamma \vdash ^{ \theta }  \ottkw{case}\  \ottnt{a} \ \ottkw{of}\ \{ \ottkw{inl}\  \ottmv{x}  \Rightarrow  \ottnt{b_{{\mathrm{1}}}}  ;
           \ottkw{inr}\  \ottmv{x}  \Rightarrow  \ottnt{b_{{\mathrm{2}}}}  \}  : \ottnt{B} }{%
{\ottdrulename{TCase}}{}%
}}

\newcommand{\ottdruleTBoxL}[1]{\ottdrule[#1]{%
\ottpremise{ \Gamma \vdash ^{ \ottkw{L} } \ottnt{a} : \ottnt{A} }%
}{
 \Gamma \vdash ^{ \ottkw{L} } \ottkw{box} \, \ottnt{a} : \ottnt{A}  \ottsym{@}  \theta }{%
{\ottdrulename{TBoxL}}{}%
}}

\newcommand{\ottdruleTBoxLV}[1]{\ottdrule[#1]{%
\ottpremise{ \Gamma \vdash ^{ \ottkw{P} } \ottnt{v} : \ottnt{A} }%
}{
 \Gamma \vdash ^{ \ottkw{L} } \ottkw{box} \, \ottnt{v} : \ottnt{A}  \ottsym{@}  \ottkw{P} }{%
{\ottdrulename{TBoxLV}}{}%
}}

\newcommand{\ottdruleTBoxP}[1]{\ottdrule[#1]{%
\ottpremise{ \Gamma \vdash ^{ \theta } \ottnt{a} : \ottnt{A} }%
}{
 \Gamma \vdash ^{ \ottkw{P} } \ottkw{box} \, \ottnt{a} : \ottnt{A}  \ottsym{@}  \theta }{%
{\ottdrulename{TBoxP}}{}%
}}

\newcommand{\ottdruleTUnbox}[1]{\ottdrule[#1]{%
\ottpremise{ \Gamma \vdash ^{ \theta } \ottnt{a} : \ottnt{A}  \ottsym{@}  \theta' }%
\ottpremise{  \Gamma ,  \ottmv{x}  :^{ \theta' }  \ottnt{A}  \vdash ^{ \theta } \ottnt{b} : \ottnt{B} }%
}{
 \Gamma \vdash ^{ \theta } \ottkw{unbox} \, \ottmv{x}  \ottsym{=}  \ottnt{a} \, \ottkw{in} \, \ottnt{b} : \ottnt{B} }{%
{\ottdrulename{TUnbox}}{}%
}}

\newcommand{\ottdruleTRoll}[1]{\ottdrule[#1]{%
\ottpremise{ \Gamma \vdash ^{ \ottkw{P} } \ottnt{a} : \ottsym{[}  \mu \, \alpha  \ottsym{.}  \ottnt{A}  \ottsym{/}  \alpha  \ottsym{]}  \ottnt{A} }%
}{
 \Gamma \vdash ^{ \ottkw{P} } \ottkw{roll} \, \ottnt{a} : \mu \, \alpha  \ottsym{.}  \ottnt{A} }{%
{\ottdrulename{TRoll}}{}%
}}

\newcommand{\ottdruleTUnroll}[1]{\ottdrule[#1]{%
\ottpremise{ \Gamma \vdash ^{ \ottkw{P} } \ottnt{a} : \mu \, \alpha  \ottsym{.}  \ottnt{A} }%
}{
 \Gamma \vdash ^{ \ottkw{P} } \ottkw{unroll} \, \ottnt{a} : \ottsym{[}  \mu \, \alpha  \ottsym{.}  \ottnt{A}  \ottsym{/}  \alpha  \ottsym{]}  \ottnt{A} }{%
{\ottdrulename{TUnroll}}{}%
}}
